\documentclass[a4paper]{JHEP3}
\usepackage{amssymb}
\usepackage{epsfig}

\title{Dynamical Generation of Yukawa Interactions
       in Intersecting D-brane Models}
\author{
Noriaki Kitazawa\\
Department of Physics, Tokyo Metropolitan University,
Hachioji, Tokyo 192-0397, Japan\\
E-mail: \email{kitazawa@phys.metro-u.ac.jp}}
\abstract{
We construct a supersymmetric composite model
 in type IIA ${\bf T^6}/({\bf Z_2} \times {\bf Z_2})$ orientifold
 with intersecting D6-branes.
Four generations of quarks and leptons
 are naturally emerged as composite fields at low energies.
Two pairs of light electroweak Higgs doublets
 are also naturally obtained.
The hierarchical Yukawa couplings
 for the quark-lepton masses can be generated 
 by the interplay between the string-level higher dimensional
 interactions among ``preons'' and
 the dynamics of the confinement of ``preons''.
Besides having four generations of quarks and leptons,
 the model is not realistic in some points:
 some exotic particles,
 one additional U$(1)$ gauge symmetry,
 no explicit mechanism for supersymmetry breaking,
 and so on.
This model is a toy model to illustrate a new mechanism
 of dynamical generation of Yukawa couplings
 for the masses and mixings of quarks and leptons.
}
\keywords{D-branes, Compactification and String Models, Quark Masses and SM Parameters, Technicolor and Composite Models}

\begin{document}

\section{Introduction}
\label{sec:intro}

The understanding of the masses and flavor mixings
 of quarks and leptons
 is one of the most important and difficult problems
 in particle physics.
In the standard model
 these masses and mixings
 are described by Yukawa coupling constants
 which are free parameters in the model.
In the field theory models beyond the standard model
 there are mainly two directions to derive these free parameters.
One direction is the unification of Yukawa coupling constants
 in grand unified theories.
The number of parameters decreases by the unification,
 and the hierarchical structure of Yukawa interactions
 is expected to be generated through the dynamics
 of spontaneous breaking of the unified gauge symmetry.
Another direction is
 the compositeness of Higgs and/or quarks and leptons.
The introduction of compositeness makes impossible
 to have Yukawa interactions in renormalizable way
 at the tree level,
 and the hierarchical Yukawa interactions
 are expected to be generated through the dynamics of
 strong coupling gauge interactions.

In string models we can expect that
 the Yukawa interactions for the masses of quarks and leptons
 are directly generated at the string level.
The main difficulty of string models is to find the vacuum
 on which the particle spectrum of the standard model is realized
 without light exotic particles.
Recent developments
 in the string models based on intersecting D-branes
 (see refs.\cite{BGKL,AFIRU-1,AFIRU-2,CSU} for the essential idea)
 suggest a way to obtain realistic models.
Especially,
 the models with low-energy supersymmetry
 \cite{CSU,BGO,Honecker,Larosa-Pradisi,
       Cvetic-Papadimitriou,kitazawa}
 are interesting,
 because these models are constructed as stable solutions
 of the perturbative string theory.
Especially for the model of ref.\cite{CSU},
 two important properties,
 the Yukawa interactions for quark-lepton masses\cite{CLS}
 and supersymmetry breaking\cite{CLW},
 are concretely discussed.
In this paper we propose
 a new mechanism to generate Yukawa interactions
 for quark-lepton masses by explicitly constructing a string model.

Consider the type IIA superstring theory compactified on
 ${\bf T}^6/({\bf Z}_2 \times {\bf Z}_2)$ orientifold, where
 ${\bf T^6}={\bf T^2} \times {\bf T^2} \times {\bf T^2}$.
The type IIA theory is invariant
 under the ${\bf Z_2} \times {\bf Z_2}$ transformation
\begin{eqnarray}
 \theta: && \qquad X_{\pm}^k \rightarrow e^{\pm i 2\pi v_k} X_{\pm}^k,
\\
 \omega: && \qquad X_{\pm}^k \rightarrow e^{\pm i 2\pi w_k} X_{\pm}^k,
\end{eqnarray}
 where $v=(0,0,1/2,-1/2,0)$ and $w=(0,0,0,1/2,-1/2)$ and
\begin{equation}
 X_{\pm}^k = 
  \left\{
   \begin{array}{ll}
    {1 \over \sqrt{2}} \left( \pm X^{2k} + X^{2k+1} \right),
     & \quad \mbox{for $k=0$}, \\
    {1 \over \sqrt{2}} \left( X^{2k} \pm i X^{2k+1} \right),
     & \quad \mbox{for $k=1,2,3,4$}
   \end{array}
  \right.
\end{equation}
 with space-time coordinates $X^\mu$, $\mu=0,1,\cdots,9$.
The type IIA theory
 is also invariant under the $\Omega R$ transformation,
 where $\Omega$ is the world-sheet parity transformation and
\begin{equation}
 R: \qquad\quad
  \left\{
   \begin{array}{ll}
    X^i \rightarrow X^i,
     & \quad \mbox{for $i=0,1,2,3,4,6,8$}, \\
    X^j \rightarrow -X^j,
     & \quad \mbox{for $j=5,7,9$}.
   \end{array}
  \right.
\end{equation}
We mod out the theory
 by the actions of $\theta$, $\omega$, $\Omega R$
 and their independent combinations.

A D6${}_a$-brane stretching over our three-dimensional space
 and winding in compact
 ${\bf T^2} \times {\bf T^2} \times {\bf T^2}$ space
 is specified by the winding numbers in each torus:
\begin{equation}
 [(n_a^1, m_a^1), (n_a^2, m_a^2), (n_a^3, m_a^3)].
\end{equation}
A D6${}_a$-brane is always accompanied by its orientifold image
 D6${}_{a'}$ whose winding numbers are
\begin{equation}
 [(n_a^1, -m_a^1), (n_a^2, -m_a^2), (n_a^3, -m_a^3)].
\end{equation}
The number of intersection
 between D6${}_a$-brane and D6${}_b$-brane is given by
\begin{equation}
 I_{ab} = \prod_{i=1}^3 \left( n_a^i m_b^i - m_a^i n_b^i \right).
\end{equation}
The intersecting angles, $\theta_a^i$,
 in each torus between D6${}_a$-brane and
 $X^4$, $X^6$ and $X^8$ axes are given by
\begin{equation}
 \theta_a^i = \tan^{-1} \left( \chi_i {{m_a^i} \over {n_a^i}} \right),
\end{equation}
 where $\chi_i$ are the ratios of two radii of each torus:
 $\chi_i \equiv R^{(i)}_2 / R^{(i)}_1$.
The system has supersymmetry,
 if $\theta_a^1+\theta_a^2+\theta_a^3=0$ is satisfied for all $a$.
The configuration of intersecting D6-branes should satisfy
 the following Ramond-Ramond tadpole cancellation conditions.
\begin{eqnarray}
 \sum_a N_a n_a^1 n_a^2 n_a^3 &=& 16,
\\
 \sum_a N_a n_a^1 m_a^2 m_a^3 &=& -16,
\\
 \sum_a N_a m_a^1 n_a^2 m_a^3 &=& -16,
\\
 \sum_a N_a m_a^1 m_a^2 n_a^3 &=& -16,
\end{eqnarray}
 where $N_a$ is the multiplicity of D6${}_a$-brane,
 and we are assuming three rectangular (untilted) tori.
The Neveu-Schwarz-Neveu-Schwarz tadpoles are automatically cancelled,
 when both the Ramond-Ramond tadpole cancellation conditions
 and the supersymmetry conditions are satisfied.

There are four sectors of open string
 depending on the D6-branes on which 
 two ends of open string are fixed:
 $aa$, $ab+ba$, $ab'+b'a$, $aa'+a'a$ sectors.
Each sector gives matter fields
 in low-energy four-dimensional space-time.
The general massless field contents
 are given in table \ref{general-spectrum}.
\TABLE{
 \begin{tabular}{|c|l|}
  \hline
  sector & field \\
  \hline\hline
  $aa$   & U$(N_a/2)$ or USp$(N_a)$ gauge multiplet. \\
         & 3 U$(N_a/2)$ adjoint or 3 USp$(N_a)$ anti-symmetric tensor
           chiral multiplets. \\
  \hline
  $ab+ba$ & $I_{ab}$ $( \Box_a, {\bar \Box}_b )$ chiral multiplets. \\
  \hline
  $ab'+b'a$ & $I_{ab'}$ $( \Box_a, \Box_b )$ chiral multiplets. \\
  \hline
  $aa'+a'a$ & ${1 \over 2}
                \left( I_{aa'} - {4 \over {2^k}} I_{aO6} \right)$
                symmetric tensor chiral multiplets. \\
            & ${1 \over 2}
                \left( I_{aa'} + {4 \over {2^k}} I_{aO6} \right)$
                anti-symmetric tensor chiral multiplets. \\
  \hline
 \end{tabular}
\caption{
General massless field contents on intersecting D6-branes.
In $aa$ sector,
 the gauge symmetry is USp$(N_a)$ or U$(N_a/2)$ corresponding to
 whether D6${}_a$-brane is parallel or not
 to a certain O6-plane, respectively.
In $aa'+a'a$ sector, $k$ is the number of tilted torus,
 and $I_{aO6}$ is the sum of the intersection numbers
 between D6${}_a$-brane and all O6-planes.
}
\label{general-spectrum}
} 
A common problem of the model building in this framework
 is the appearance of the massless adjoint fields in $aa$ sector,
 since there are no massless matter field
 in the adjoint representation
 under the standard model gauge group in Nature.
These fields are expected to be massive
 in case of the appropriate curved compact space,
 because these are the moduli fields of D6-brane configurations.

In the next section
 we construct a supersymmetric composite model.
The standard model gauge groups are originated from the D6-branes
 which are not parallel to any O6-planes.
Tadpole cancellation conditions can be satisfied
 by introducing several additional D6-branes
 which are parallel to certain O6-planes.
Since the USp$(N)$ gauge interactions on these D6-branes
 are stronger than the U$(N)$ gauge interactions in general,
 it is natural to identify them to the confining dynamics
 among ``preons''.
Four generations of quarks and leptons are naturally emerged.
Two pairs of light electroweak Higgs doublets
 are also naturally obtained.
In section \ref{sec:yukawa}
 the dynamical generation of
 Yukawa interactions for the quark-lepton mass and mixing
 is discussed.
The hierarchical Yukawa coupling matrices can be generated 
 by the interplay between the string-level higher dimensional
 interactions among ``preons'' and
 the dynamics of the confinement of ``preons''.
We are not going to calculate
 all the Yukawa coupling constants in this toy model,
 but instead we give a general scenario
 to have small Yukawa coupling constants
 and describe an example of having quark-lepton mass differences.
In the last section we present a summary and conclusions.

\section{The Model and Low-energy Dynamics}
\label{sec:model}

The D6-brane configuration of the model
 is given in table \ref{config}.
\TABLE{
 \begin{tabular}{|c|c|c|}
  \hline
  D6-brane & winding number & multiplicity     \\
  \hline\hline
  D6${}_1$   & $ \quad [(1,-1), (1,1), (1,0)] \quad $ & $4$ \\
  \hline
  D6${}_2$   & $ \quad [(1,1), (1,0), (1,-1)] \quad $ & $6+2$  \\
  \hline
  D6${}_3$   & $ \quad [(1,0), (1,-1), (1,1)] \quad $  & $2+2$  \\
  \hline
  D6${}_4$   & $ \quad [(1,0), (0,1), (0,-1)] \quad $ & $12$  \\
  \hline
  D6${}_5$   & $ \quad [(0,1), (1,0), (0,-1)] \quad $ & $8$ \\
  \hline
  D6${}_6$   & $ \quad [(0,1), (0,-1), (1,0)] \quad $ & $12$ \\
  \hline
 \end{tabular}
\caption{
Configuration of intersecting D6-branes.
All three tori are considered to be rectangular (untilted).
Three D6-branes, D6${}_4$, D6${}_5$ and D6${}_6$,
 are on top of certain O6-planes.
}
\label{config}
}
Both tadpole cancellation conditions and supersymmetry conditions
 are satisfied in this configuration under the condition of
 $\chi_1 = \chi_2 = \chi_3 \equiv \chi$.
D6${}_2$-brane consists of
 two parallel D6-branes of multiplicity six and two
 which are separated in the second torus
 in a consistent way with the orientifold projections.
D6${}_3$-brane consists of
 two parallel D6-branes of multiplicity two
 which are separated in the first torus
 in a consistent way with the orientifold projections.
D6${}_1$, D6${}_2$ and D6${}_3$ branes give gauge symmetries
 of U$(2)_L=$SU$(2)_L \times$U$(1)_L$,
 U$(3)_c \times$U$(1) =$SU$(3)_c \times$U$(1)_c \times$U$(1)$
 and U$(1)_1 \times$U$(1)_2$,
 respectively.
The hypercharge is defined as
\begin{equation}
 {Y \over 2} = {1 \over 2} \left( {{Q_c} \over 3} - Q \right)
             + {1 \over 2} \left( Q_1 - Q_2 \right),
\end{equation}
 where $Q_c$, $Q$, $Q_1$ and $Q_2$ are charges of
 U$(1)_c$, U$(1)$, U$(1)_1$ and U$(1)_2$, respectively.
There is another non-anomalous U(1) gauge symmetry,
 U$(1)_R$, whose charge is defined as
\begin{equation}
 Q_R = Q_1 - Q_2.
\end{equation}
The remaining three U$(1)$ gauge symmetries
 which are generated by
 $Q_L$ (namely U$(1)_L$), $Q_c+Q$ and $Q_1+Q_2$ are anomalous,
 and their gauge bosons have masses of the order of the string scale.
These three anomalous U$(1)$ gauge symmetries
 are independent from the two non-anomalous U$(1)$ gauge symmetries:
 ${\rm tr}((Y/2) Q_L) = 0$, for example.

D6${}_4$-brane,
 which consists of twelve D6-branes,
 gives the gauge symmetry of USp$(12)_{{\rm D6}_4}$,
 if all the twelve D6-branes are on top of one O6-plane.
Since there are eight O6-planes with the same winding numbers,
 there are some ambiguities to set the place of twelve D6-branes,
 and the actual gauge symmetry is the subgroup of
 USp$(12)_{{\rm D6}_4}$ in general.
Figure \ref{usp} gives
 one configuration which we take in this paper.
The resultant gauge symmetry is
\begin{equation}
 \mbox{USp}(2)_{{\rm D6}_4,1} \times
 \mbox{USp}(2)_{{\rm D6}_4,2} \times
 \mbox{USp}(2)_{{\rm D6}_4,3} \times
 \mbox{USp}(2)_{{\rm D6}_4,4} \times
 \mbox{USp}(2)_{{\rm D6}_4,5} \times
 \mbox{USp}(2)_{{\rm D6}_4,6}.
\label{gauge_D6_4}
\end{equation}
This is the same for D6${}_5$ and D6${}_6$ branes.
The resultant gauge symmetries
 by D6${}_5$ and D6${}_6$ branes are
\begin{equation}
 \mbox{USp}(2)_{{\rm D6}_5,1} \times
 \mbox{USp}(2)_{{\rm D6}_5,2} \times
 \mbox{USp}(2)_{{\rm D6}_5,3} \times
 \mbox{USp}(2)_{{\rm D6}_5,4}
\label{gauge_D6_5}
\end{equation}
 and
\begin{equation}
 \mbox{USp}(2)_{{\rm D6}_6,1} \times
 \mbox{USp}(2)_{{\rm D6}_6,2} \times
 \mbox{USp}(2)_{{\rm D6}_6,3} \times
 \mbox{USp}(2)_{{\rm D6}_6,4} \times
 \mbox{USp}(2)_{{\rm D6}_6,5} \times
 \mbox{USp}(2)_{{\rm D6}_6,6},
\label{gauge_D6_6}
\end{equation}
 respectively.
\EPSFIGURE{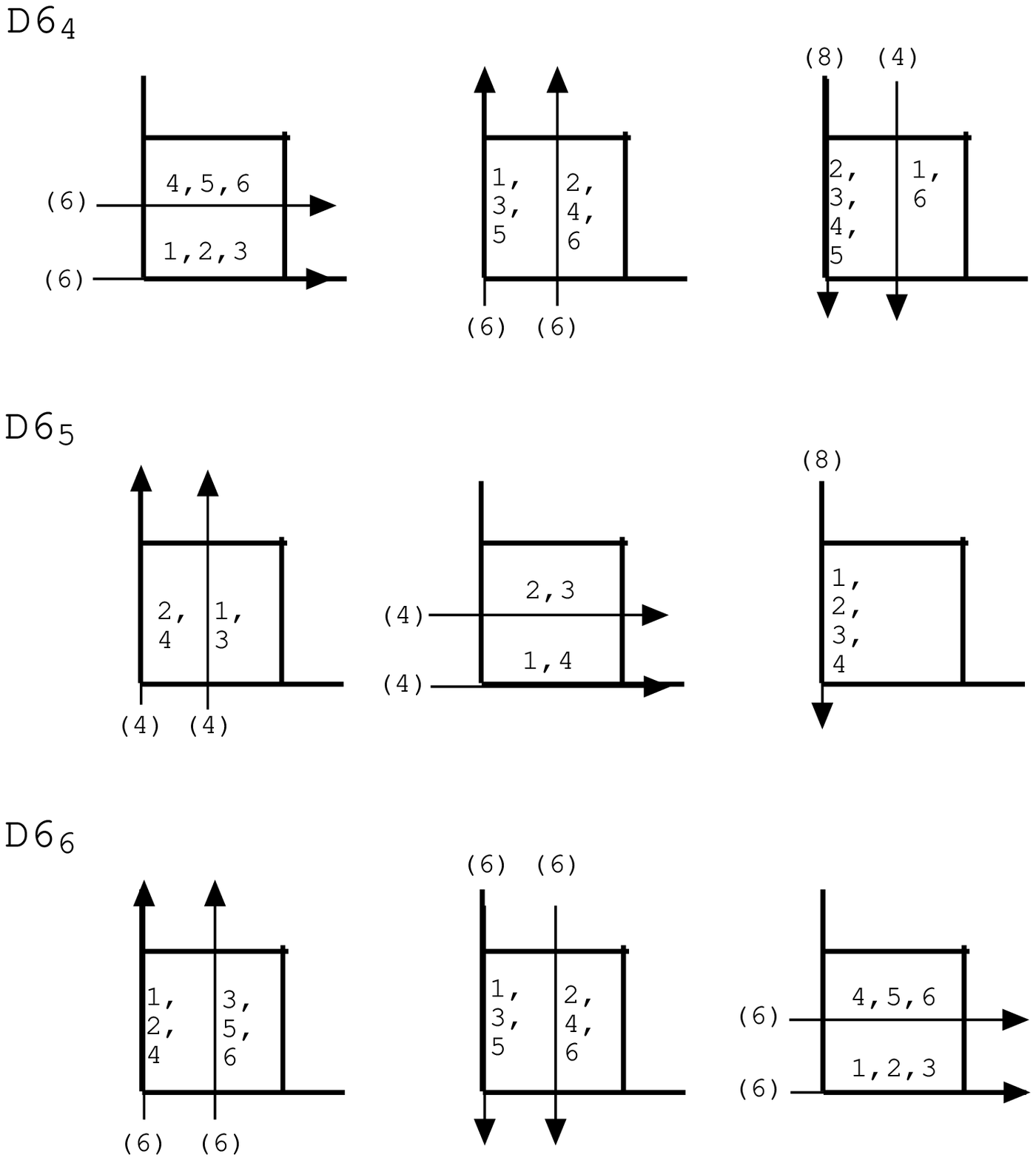,width=11cm}
{\label{usp}
Configurations of twelve, eight and twelve D6-branes
 of D6${}_4$, D6${}_5$ and D6${}_6$ branes, respectively.
The numbers in brackets are multiplicities of D6-brane stacks,
 and the numbers without brackets specify
 one of the USp$(2)$ gauge groups in eqs. (\ref{gauge_D6_4}),
 (\ref{gauge_D6_5}) and (\ref{gauge_D6_6}). 
}

The value of the coupling constant of U$(N)$ gauge interaction
 at the string scale $M_s$ is determined by
\begin{equation}
 g_{\rm U}^2 = \sqrt{4\pi} \kappa_4 M_s {\sqrt{V_6} \over {V_3}},
\end{equation}
 where $\kappa_4=\sqrt{8 \pi G_N}$, $M_s = 1 / \sqrt{\alpha'}$,
 $V_6$ is the volume of compact six-dimensional space
 and $V_3$ is the volume of corresponding D6-brane
 in compact six-dimensional space\cite{CLW}.
In case of that the D6-brane is parallel to a certain O6-plane,
 the corresponding USp$(N)$ gauge coupling constant is given by
\begin{equation}
 g_{\rm USp}^2 = 2 \sqrt{4\pi} \kappa_4 M_s {\sqrt{V_6} \over {V_3}},
\end{equation}
 which has an extra factor $2$\cite{BLS}.
In our model, all the USp$(2)$ gauge interactions of
 eqs. (\ref{gauge_D6_4}), (\ref{gauge_D6_5}) and (\ref{gauge_D6_6})
 naturally have larger gauge coupling constants than
 the gauge coupling constants of any other gauge interactions,
 because of smaller $V_3$ and the extra factor $2$.
We call
 these USp$(2)$ gauge interactions ``hypercolor'' interactions.
These ``hypercolor'' interactions
 give the confining force of ``preons''.
The actual values of
 the gauge coupling constants at the string scale in our model
 are given by
\begin{eqnarray}
 \alpha_{\rm U} &=& {2 \over {\sqrt{4\pi}}} \kappa_4 M_s
                    {{\chi^{3/2}} \over {1+\chi^2}},
\\
 \alpha_{\rm USp} &=& {4 \over {\sqrt{4\pi}}} \kappa_4 M_s
                      {1 \over \sqrt{\chi}}
\end{eqnarray}
 for all U$(N)$ and USp$(N)$ type gauge groups, respectively.
We have a relation
 between three gauge couplings of the standard model
 at the string scale as
\begin{equation}
 \alpha_3 = \alpha_2 = {7 \over 9} \alpha_Y.
\end{equation}
The value of
 the U$(1)_R$ gauge coupling constant at the string scale
 is given by $\alpha_R = {7 \over {18}} \alpha_Y$.
If we choose $\kappa_4 M_s \sim 1$ and $\chi \sim 0.1$,
 the values of the scales of dynamics of all USp$(2)$
 gauge interactions are of the order of $M_s$,
 and the values of the standard model gauge coupling constants
 are reasonably of the order of $1/100$ at the string scale.

A schematic picture
 of the configuration of intersecting D6-branes of this model
 is given in figure \ref{intersec}.
\EPSFIGURE{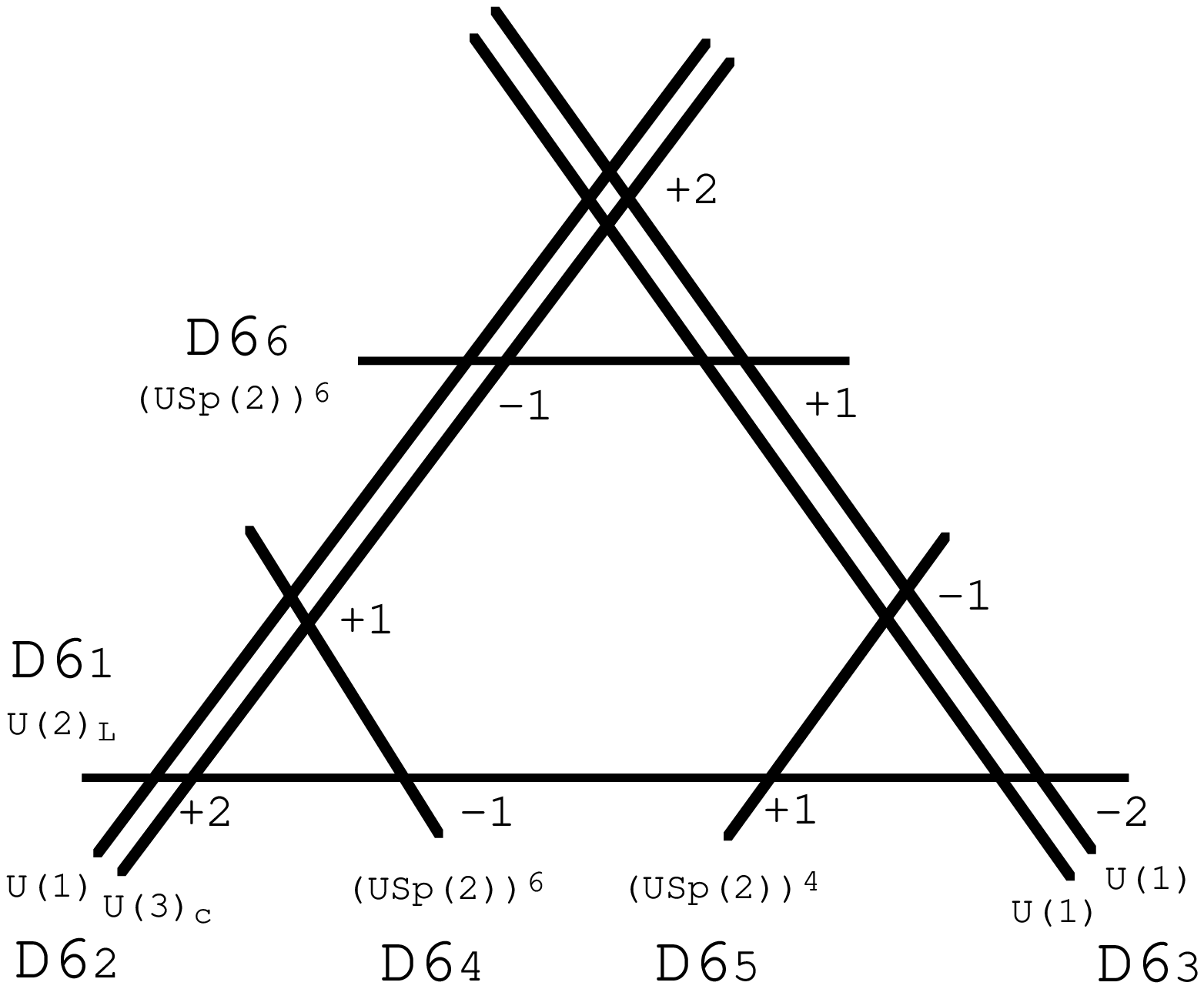,height=100mm}
{\label{intersec}
Schematic picture of the configuration of intersecting D6-branes.
This picture describes
 only the situation of the intersection of D6-branes,
 and the relative place of each D6-brane has no meaning.
The number at the intersection point
 between D6${}_a$ and D6${}_b$ branes
 denotes intersection number $I_{ab}$ with $a<b$.
}
There are no $ab'+b'a$, $aa'+a'a$ sectors of open string
 in this configuration.
The massless particle contents are given in table \ref{contents_1}.
\TABLE{
 \begin{tabular}{|c|c|c|}
  \hline
  sector             & $\mbox{SU}(3)_c \times \mbox{SU}(2)_L
                       \times \mbox{USp}(8)
                       \times \mbox{USp}(12)_{D6_4}
                       \times \mbox{USp}(12)_{D6_6}$
                     & field  \\
                     & ($Y/2, Q_R$)($Q_L, Q_c+Q, Q_1+Q_2$)
                     &        \\
  \hline\hline
  $D6_1 \cdot D6_2$  & $(3^*, 2, 1, 1, 1)_{(-1/6,0)(+1,-1,0)}
                         \times 2$
                     & ${\bar q}_i$ \\
                     & $(1, 2, 1, 1, 1)_{(+1/2,0)(+1,-1,0)}
                         \times 2$
                     & ${\bar l}_i$  \\
  \hline
  $D6_1 \cdot D6_4$  & $(1, 2, 1, 12, 1)_{(0,0)(-1,0,0)}$
                     & $D$ \\
  \hline
  $D6_2 \cdot D6_4$  & $(3, 1, 1, 12, 1)_{(+1/6,0)(0,+1,0)}$
                     & $C$  \\
                     & $(1, 1, 1, 12, 1)_{(-1/2,0)(0,+1,0)}$
                     & $N$  \\
  \hline
  $D6_1 \cdot D6_3$  & $(1, 2, 1, 1, 1)_{(+1/2,+1)(-1,0,+1)}
                        \times 2$
                     & $H^{(1)}_i$ \\
                     & $(1, 2, 1, 1, 1)_{(-1/2,-1)(-1,0,+1)}
                        \times 2$
                     & ${\bar H}^{(2)}_i$ \\
  \hline
  $D6_1 \cdot D6_5$  & $(1, 2, 8, 1, 1)_{(0,0)(+1,0,0)}$
                     & $T$ \\
  \hline
  $D6_3 \cdot D6_5$  & $(1, 1, 8, 1, 1)_{(+1/2,+1)(0,0,-1)}$
                     & $T^{(+)}$ \\
                     & $(1, 1, 8, 1, 1)_{(-1/2,-1)(0,0,-1)}$
                     & $T^{(-)}$ \\
  \hline
  $D6_2 \cdot D6_3$  & $(3, 1, 1, 1, 1)_{(-1/3,-1)(0,+1,-1)}
                        \times 2$
                     & ${\bar d}_i$ \\
                     & $(3, 1, 1, 1, 1)_{(+2/3,+1)(0,+1,-1)}
                        \times 2$
                     & ${\bar u}_i$ \\
                     & $(1, 1, 1, 1, 1)_{(-1,-1)(0,+1,-1)}
                        \times 2$
                     & ${\bar e}_i$ \\
                     & $(1, 1, 1, 1, 1)_{(0,+1)(0,+1,-1)}
                        \times 2$
                     & ${\bar \nu}_i$ \\
  \hline
  $D6_2 \cdot D6_6$  & $(3^*, 1, 1, 1, 12)_{(-1/6,0)(0,-1,0)}$
                     & ${\bar C}$ \\
                     & $(1, 1, 1, 1, 12)_{(+1/2,0)(0,-1,0)}$
                     & ${\bar N}$ \\
  \hline
  $D6_3 \cdot D6_6$  & $(1, 1, 1, 1, 12)_{(+1/2,+1)(0,0,+1)}$
                     & ${\bar D}^{(+)}$ \\
                     & $(1, 1, 1, 1, 12)_{(-1/2,-1)(0,0,+1)}$
                     & ${\bar D}^{(-)}$ \\
  \hline
 \end{tabular}
\caption{
Low-energy particle contents before ``hypercolor'' confinement.
Here, we use the representations of USp$(12)$ and USp$(8)$ groups
 instead of using the representations of many USp$(2)$ groups,
 for simplicity.
The fields from $aa$ sectors are neglected for simplicity.
}
\label{contents_1}
}

From figure \ref{intersec}
 we see that there are the following Yukawa interactions
 at the string level.
\begin{eqnarray}
 W_{\rm left} &=& {\bar q}_i C D + {\bar l}_i N D,
\label{yukawa_left}
\\
 W_{\rm right} &=& {\bar u}_i {\bar C} {\bar D}^{(-)}
                 + {\bar d}_i {\bar C} {\bar D}^{(+)}
                 + {\bar \nu}_i {\bar N} {\bar D}^{(-)}
                 + {\bar e}_i {\bar N} {\bar D}^{(+)},
\label{yukawa_right}
\\
 W_{\rm higgs} &=& H_i^{(1)} T T^{(-)} + {\bar H}_i^{(2)} T T^{(+)},
\label{yukawa_higgs}
\\
 W_{\rm massive} &=& {\bar q}_i {\bar u}_j {\bar H}_k^{(2)}
                   + {\bar q}_i {\bar d}_j H_k^{(1)}
                   + {\bar l}_i {\bar \nu}_j {\bar H}_k^{(2)}
                   + {\bar l}_i {\bar e}_j H_k^{(1)},
\end{eqnarray}
 where $i,j,k=1,2$.
Here,
 we just notice the existence of these Yukawa interactions
 without specifying the value of their coupling constants.
The Yukawa interactions of
 eqs. (\ref{yukawa_left}) and (\ref{yukawa_right}) 
 are the necessary ingredients to give masses to some exotic fields.

We also see from figure \ref{intersec} the existence of
 the following higher dimensional interactions at the string level.
\begin{equation}
 W_{\rm yukawa} = C D {\bar C} {\bar D}^{(-)} T T^{(+)}
                + C D {\bar C} {\bar D}^{(+)} T T^{(-)}
                + N D {\bar N} {\bar D}^{(-)} T T^{(+)}
                + N D {\bar N} {\bar D}^{(+)} T T^{(-)}.
\label{to_be_yukawa}
\end{equation}
Here, we also just notice their existence
 without specifying the value of their coupling constants.
These interactions
 give Yukawa interactions for the quark-lepton mass
 after the ``hypercolor'' confinement at low energies.

In the following,
 we call the sectors of
 D6${}_1$-D6${}_2$-D6${}_4$,
 D6${}_2$-D6${}_3$-D6${}_6$ and
 D6${}_1$-D6${}_3$-D6${}_5$,
 as left-handed, right-handed and Higgs sectors, respectively.
Each sector gives
 left-handed quarks and leptons,
 right-handed quarks and leptons
 and Higgs doublets, respectively.
We explain the result of the confinement
 by the strong USp$(2)$ ``hypercolor'' dynamics
 in each sector in order in the following.

\subsection{Dynamics of the Left-handed Sector}
\label{subsec:left}

The dynamics in this sector
 is very similar to the one in the model of ref.\cite{kitazawa}.
The field contents of this sector
 are given in table \ref{contents_left}.
\TABLE{
 \begin{tabular}{|c|c|c|}
  \hline
  sector             & $\left(
                         \mbox{SU}(3)_c \times \mbox{SU}(2)_L
                        \right)
                        \times
                        \left(
                         \mbox{USp}(2)_1 \times
                         \mbox{USp}(2)_2 \times
                         \mbox{USp}(2)_3 \times
                         \mbox{USp}(2)_4
                        \right.$
                     & field \\
                     & $\left.
                         \times
                         \mbox{USp}(2)_5 \times                  
                         \mbox{USp}(2)_6                         
                        \right)_{D6_4}$
                        ($Y/2, Q_R$)($Q_L, Q_c+Q, Q_1+Q_2$)
                     & \\
  \hline\hline
  $D6_1 \cdot D6_2$  & $(3^*, 2)(1,1,1,1,1,1)_{(-1/6,0)(+1,-1,0)}
                         \times 2$
                     & ${\bar q}_i$ \\
                     & $(1, 2)(1,1,1,1,1,1)_{(+1/2,0)(+1,-1,0)}
                         \times 2$
                     & ${\bar l}_i$  \\
  \hline
  $D6_1 \cdot D6_4$  & $(1, 2)(2,1,1,1,1,1)_{(0,0)(-1,0,0)}$
                     & $D_\alpha$ \\
                     & $(1, 2)(1,2,1,1,1,1)_{(0,0)(-1,0,0)}$
                     & \\
                     & $(1, 2)(1,1,2,1,1,1)_{(0,0)(-1,0,0)}$
                     & \\
                     & $(1, 2)(1,1,1,2,1,1)_{(0,0)(-1,0,0)}$
                     & \\
                     & $(1, 2)(1,1,1,1,2,1)_{(0,0)(-1,0,0)}$
                     & \\
                     & $(1, 2)(1,1,1,1,1,2)_{(0,0)(-1,0,0)}$
                     & \\
  \hline
  $D6_2 \cdot D6_4$  & $(3, 1)(2,1,1,1,1,1)_{(+1/6,0)(0,+1,0)}$
                     & $C_\alpha$  \\
                     & $(3, 1)(1,2,1,1,1,1)_{(+1/6,0)(0,+1,0)}$
                     & \\
                     & $(3, 1)(1,1,2,1,1,1)_{(+1/6,0)(0,+1,0)}$
                     & \\
                     & $(3, 1)(1,1,1,2,1,1)_{(+1/6,0)(0,+1,0)}$
                     & \\
                     & $(3, 1)(1,1,1,1,2,1)_{(+1/6,0)(0,+1,0)}$
                     & \\
                     & $(3, 1)(1,1,1,1,1,2)_{(+1/6,0)(0,+1,0)}$
                     & \\
                     & $(1, 1)(2,1,1,1,1,1)_{(-1/2,0)(0,+1,0)}$
                     & $N_\alpha$  \\
                     & $(1, 1)(1,2,1,1,1,1)_{(-1/2,0)(0,+1,0)}$
                     & \\
                     & $(1, 1)(1,1,2,1,1,1)_{(-1/2,0)(0,+1,0)}$
                     & \\
                     & $(1, 1)(1,1,1,2,1,1)_{(-1/2,0)(0,+1,0)}$
                     & \\
                     & $(1, 1)(1,1,1,1,2,1)_{(-1/2,0)(0,+1,0)}$
                     & \\
                     & $(1, 1)(1,1,1,1,1,2)_{(-1/2,0)(0,+1,0)}$
                     & \\
  \hline
 \end{tabular}
\caption{
Field contents of the left-handed sector.
Here, $i=1,2$ and $\alpha = 1,2, \cdots ,6$.
}
\label{contents_left}
}
For USp$(2)_1$ gauge interaction, for example,
 there are three massless flavors,
 and the low-energy effective fields
 after the confinement are as follows
 (see ref.\cite{Intriligator-Pouliot} for the dynamics of
  the supersymmetric USp gauge theory).
\begin{equation}
 M_{L,1}
   = \left[
      \left(
       \begin{array}{c}
        C_1 \\ D_1 \\ N_1
       \end{array}
      \right)
      \left(
       \begin{array}{ccc}
        C_1 & D_1 & N_1
       \end{array}
      \right)
          \right]
   = \left(
      \begin{array}{ccc}
       [C_1 C_1] & [C_1 D_1] & [C_1 N_1] \\
                 & [D_1 D_1] & [D_1 N_1] \\
                 &           & [N_1 N_1]
      \end{array}
     \right),
\end{equation}
 where square brackets denote the anti-symmetric contraction
 of the indices of the fundamental representation of USp$(2)_1$.
The field contents after the ``hypercolor'' confinement
 are given in table \ref{contents_left_confined}.
\TABLE{
 \begin{tabular}{|c|c|}
  \hline
  field             & $\mbox{SU}(3)_c \times \mbox{SU}(2)_L$ \\
                    & ($Y/2, Q_R$)($Q_L, Q_c+Q, Q_1+Q_2$) \\
  \hline\hline
  ${\bar q}_i$      & $(3^*, 2)_{(-1/6,0)(+1,-1,0)}
                       \times 2$ \\
  ${\bar l}_i$      & $(1, 2)_{(+1/2,0)(+1,-1,0)}
                       \times 2$ \\
  \hline
  $[C_\alpha C_\alpha] \sim {\bar d}'_\alpha$
                    & $(3^*, 1)_{(+1/3,0)(0,+2,0)}
                       \times 6$ \\
  $[C_\alpha D_\alpha] \sim q_\alpha$
                    & $(3, 2)_{(+1/6,0)(-1,+1,0)}
                       \times 6$ \\
  $[C_\alpha N_\alpha] \sim \Phi_\alpha$
                    & $(3, 1)_{(-1/3,0)(0,+2,0)}
                       \times 6$ \\
  $[D_\alpha D_\alpha] \sim S_\alpha$
                    & $(1, 1)_{(0,0)(-2,0,0)}
                       \times 6$ \\
  $[D_\alpha N_\alpha] \sim l_\alpha$
                    & $(1, 2)_{(-1/2,0)(-1,+1,0)}
                       \times 6$ \\
  \hline
 \end{tabular}
\caption{
Field contents of the left-handed sector
 after the ``hypercolor'' confinement.
Here, $i=1,2$ and $\alpha = 1,2, \cdots ,6$.
}
\label{contents_left_confined}
}
The following superpotential among the composite fields
 is dynamically generated.
\begin{equation}
 W_{\rm dyn}^{\rm left}
 = - \sum_{\alpha = 1}^6
     {1 \over {\Lambda_{L,\alpha}}^3}
     {\rm Pf} M_{L,\alpha}
 = - \sum_{\alpha = 1}^6
     {1 \over {\Lambda_{L,\alpha}}^3}
     \left(
      {\bar d}'_\alpha q_\alpha l_\alpha
      + {\bar d}'_\alpha \Phi_\alpha S_\alpha
      + q_\alpha q_\alpha \Phi_\alpha
     \right),
\label{dynamical-W-left}    
\end{equation}
 where $\Lambda_{L,\alpha}$ is the scale of dynamics of 
 each USp$(2)_\alpha$.
Since the values of all gauge coupling constants of
 USp$(2)_\alpha$ are equal at the string scale,
 all the scales of dynamics are equal to a single $\Lambda_L$.
The dynamically generated superpotential
 is invariant under all the gauge symmetry transformations
 once the moduli field dependence of $\Lambda_L$ is considered.
The component of the Ramond-Ramond field
 which acts an important role for the anomaly cancellation
 is also the imaginary part of the moduli field
 in the scale of dynamics.

If the fields $S_\alpha$ have vacuum expectation values,
 the exotic fields of
 ${\bar d}'_\alpha$ and $\Phi_\alpha$ become massive and
 only $q_\alpha$ and $l_\alpha$ remain massless.
The expected mechanism for the vacuum expectation value
 is the emergence of the Fayet-Iliopoulos term
 to the anomalous U$(1)_L$ in some general compact space,
 blown-up orientifold\cite{IRU}, for example.
In this paper
 we simply assume large vacuum expectation values
 of all $S_\alpha$.

The Yukawa interactions of eq.(\ref{yukawa_left})
 give masses for two of six fields
 of each $q_\alpha$ and $l_\alpha$.
The values of the masses
 are determined by the values of the Yukawa coupling constants
 and the scale of dynamics $\Lambda_L$.
The actual values of the Yukawa coupling constants
 are determined by the D-brane configuration.
If we take the configuration of
 D6${}_1$ and D6${}_2$ branes as figure \ref{D6_1-D6_2},
 $q_3$ and $q_5$ become massive
 pairing with ${\bar q}_1$ and ${\bar q}_2$, respectively, and
 $l_2$ and $l_4$ become massive
 pairing with ${\bar l}_1$ and ${\bar l}_2$, respectively.
\footnote{
These one-to-one pairings are satisfied in good approximation
 when the radii of all three tori are large:
 $R^{(i)}_1, R^{(i)}_2 > \sqrt{\alpha'}$
 for all $i=1,2,3$.
When the radii of the first torus are small,
 as we will take in the next section,
 ${\bar q}_1$ becomes massive by pairing with
 a linear combination of $q_3$ and $q_5$,
 and ${\bar q}_2$ becomes massive by pairing with
 another linear combination of $q_3$ and $q_5$.
The similar occurs for ${\bar l}_1$ and ${\bar l}_2$.
}
The values of these masses are of the order of $\Lambda_L$.
Here, we are neglecting the contributions
 from the Yukawa couplings which are exponentially suppressed
 as an approximation.
As the result,
 we have four generation fields of
 left-handed quarks and left-handed leptons.
\EPSFIGURE{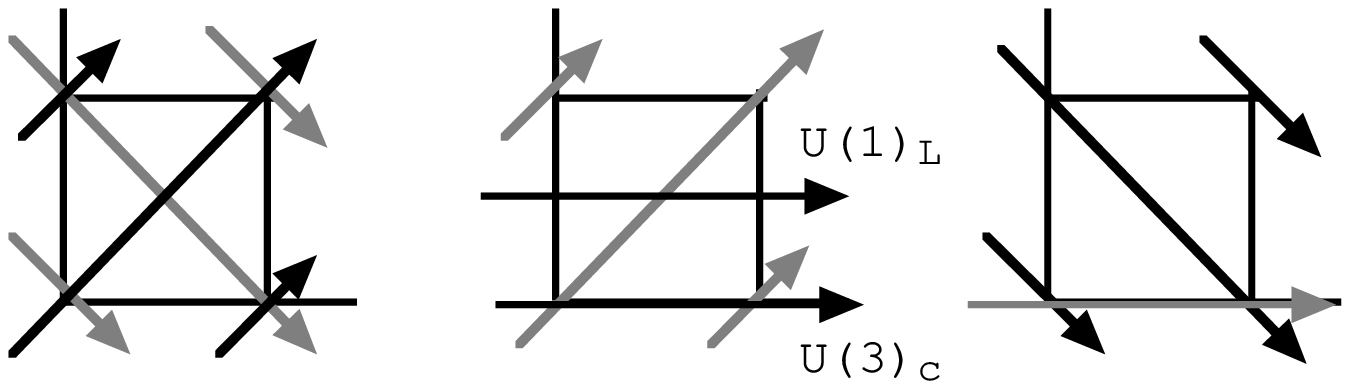,height=40mm}
{\label{D6_1-D6_2}
Actual D6-brane configuration of D6${}_1$ and D6${}_2$ branes.
Gray lines with arrow denote D6-branes of D6${}_1$
 and the solid lines with arrow denote D6-branes of D6${}_2$.
Two stacks of D6-branes of D6${}_2$ in the second torus
 result the difference between quarks and leptons.
}

\subsection{Dynamics of the Right-handed Sector}
\label{subsec:right}

Almost the same mechanism in the left-handed sector
 occurs in the right-handed sector.
Here, we are not going to describe all the details,
 but simply present the results.
The low-energy effective fields
 after the ``hypercolor'' confinement are
\begin{equation}
 M_{R,\alpha}
   = \left[
      \left(
       \begin{array}{c}
        {\bar C}_\alpha \\ {\bar N}_\alpha \\
        {\bar D}^{(+)}_\alpha \\ {\bar D}^{(-)}_\alpha
       \end{array}
      \right)
      \left(
       \begin{array}{cccc}
        {\bar C}_\alpha & {\bar N}_\alpha &
        {\bar D}^{(+)}_\alpha & {\bar D}^{(-)}_\alpha
       \end{array}
      \right)
          \right]
   = \left(
      \begin{array}{cccc}
       [{\bar C}_\alpha {\bar C}_\alpha] &
       [{\bar C}_\alpha {\bar N}_\alpha] &
       [{\bar C}_\alpha {\bar D}^{(+)}_\alpha] &
       [{\bar C}_\alpha {\bar D}^{(-)}_\alpha] \\
       &
       [{\bar N}_\alpha {\bar N}_\alpha] &
       [{\bar N}_\alpha {\bar D}^{(+)}_\alpha] &
       [{\bar N}_\alpha {\bar D}^{(-)}_\alpha] \\
       & &
       [{\bar D}^{(+)}_\alpha {\bar D}^{(+)}_\alpha] &
       [{\bar D}^{(+)}_\alpha {\bar D}^{(-)}_\alpha] \\
       & & &
       [{\bar D}^{(-)}_\alpha {\bar D}^{(-)}_\alpha]
      \end{array}
     \right),
\end{equation}
 where square brackets denote the anti-symmetric contraction
 of the indices of the fundamental representation of
 USp$(2)_\alpha$ in eq.(\ref{gauge_D6_6}).
The field contents after the ``hypercolor'' confinement
 is given in table \ref{contents_right_confined}.
\TABLE{
 \begin{tabular}{|c|c|}
  \hline
  field             & $\mbox{SU}(3)_c \times \mbox{SU}(2)_L$ \\
                    & ($Y/2, Q_R$)($Q_L, Q_c+Q, Q_1+Q_2$) \\
  \hline\hline
  ${\bar u}_i$      & $(3, 1)_{(+2/3,+1)(0,+1,-1)}
                       \times 2$ \\
  ${\bar d}_i$      & $(3, 1)_{(-1/3,-1)(0,+1,-1)}
                       \times 2$ \\
  ${\bar \nu}_i$    & $(1, 1)_{(0,+1)(0,+1,-1)}
                       \times 2$ \\
  ${\bar e}_i$      & $(1, 1)_{(-1,-1)(0,+1,-1)}
                       \times 2$ \\
  \hline
  $[{\bar C}_\alpha {\bar C}_\alpha] \sim d'_\alpha$
                    & $(3, 1)_{(-1/3,0)(0,-2,0)}
                       \times 6$ \\
  $[{\bar C}_\alpha {\bar N}_\alpha] \sim {\bar \Phi}_\alpha$
                    & $(3^*, 1)_{(+1/3,0)(0,-2,0)}
                       \times 6$ \\
  $[{\bar C}_\alpha {\bar D}^{(+)}_\alpha] \sim d_\alpha$
                    & $(3^*, 1)_{(+1/3,+1)(0,-1,+1)}
                       \times 6$ \\
  $[{\bar C}_\alpha {\bar D}^{(-)}_\alpha] \sim u_\alpha$
                    & $(3^*, 1)_{(-2/3,-1)(0,-1,+1)}
                       \times 6$ \\
  $[{\bar N}_\alpha {\bar D}^{(+)}_\alpha] \sim e_\alpha$
                    & $(1, 1)_{(+1,+1)(0,-1,+1)}
                       \times 6$ \\
  $[{\bar N}_\alpha {\bar D}^{(-)}_\alpha] \sim \nu_\alpha$
                    & $(1, 1)_{(0,-1)(0,-1,+1)}
                       \times 6$ \\
  $[{\bar D}^{(+)}_\alpha {\bar D}^{(-)}_\alpha]
                      \sim {\bar S}_\alpha$
                    & $(1, 1)_{(0,0)(0,0,+2)}
                       \times 6$ \\
  \hline
 \end{tabular}
\caption{
Field contents of the right-handed sector
 after the ``hypercolor'' confinement.
Here, $i=1,2$ and $\alpha = 1,2, \cdots ,6$.
}
\label{contents_right_confined}
}
The following superpotential among the composite fields
 is dynamically generated.
\begin{equation}
 W_{\rm dyn}^{\rm left}
 = - \sum_{\alpha = 1}^6
     {1 \over {\Lambda_{R,\alpha}}^3}
     {\rm Pf} M_{R,\alpha}
 = - \sum_{\alpha = 1}^6
     {1 \over {\Lambda_{R,\alpha}}^3}
     \left(
      d'_\alpha {\bar \Phi}_\alpha {\bar S}_\alpha
      + d'_\alpha e_\alpha u_\alpha
      + d'_\alpha \nu_\alpha d_\alpha
      + {\bar \Phi}_\alpha d_\alpha u_\alpha
     \right),
\label{dynamical-W-right}    
\end{equation}
 where $\Lambda_{R,\alpha}$ is the scale of dynamics of 
 each USp$(2)_\alpha$.
Since all the coupling constants of
 USp$(2)_\alpha$ are equal at the string scale,
 all the scales of dynamics are equal:
 $\Lambda_{R,\alpha} = \Lambda_R$.
This dynamically generated superpotential
 is invariant under all the gauge symmetry transformations
 considering the moduli dependence of $\Lambda_R$.

If all ${\bar S}_\alpha$ have vacuum expectation values,
 the fields $d'_\alpha$ and ${\bar \Phi}_\alpha$ become massive
 and only $u_\alpha$, $d_\alpha$, $\nu_\alpha$ and $e_\alpha$
 remain massless.
The expected mechanism for the vacuum expectation value
 is the emergence of the Fayet-Iliopoulos term to
 the anomalous U$(1)_{Q_1+Q_2}$ in some more general compact space.
In this paper
 we simply assume large vacuum expectation values
 of all ${\bar S}_\alpha$.
 
The Yukawa interactions of eq.(\ref{yukawa_right})
 give masses to two of six fields
 of each $u_\alpha$, $d_\alpha$, $\nu_\alpha$ and $e_\alpha$.
The values of the masses
 are determined by the values of the Yukawa coupling constants
 and the scale of dynamics $\Lambda_R$.
The actual value of the Yukawa coupling constants
 are determined by the D-brane configuration.
If we take the configuration of
 D6${}_2$ and D6${}_3$ branes as figure \ref{D6_2-D6_3},
 and take the radii of the first torus are not so large:
 $R^{(1)}_1, R^{(1)}_2 \gtrsim \sqrt{\alpha'}$,
 a linear combination of $u_1$ and $u_3$, $u_5$,
 a linear combination of $d_1$ and $d_3$, $d_5$,
 $\nu_2$,
 a linear combination of $\nu_4$ and $\nu_6$,
 $e_2$
 and a linear combination of $e_4$ and $e_6$ become massive
 by pairing with
 ${\bar u}_1$, ${\bar u}_2$, ${\bar d}_1$, ${\bar d}_2$,
 ${\bar \nu}_1$, ${\bar \nu}_2$, ${\bar e}_1$ and ${\bar e}_2$,
 respectively.
The radii of other tori
 are assumed to be large in comparison with $\sqrt{\alpha'}$,
  and we neglect small contributions from
  the exponentially suppressed Yukawa interactions
  at the string level.
The values of the masses are of the order of $\Lambda_R$.
As the result,
 we have four generation fields of
 right-handed quarks and right-handed leptons.
\EPSFIGURE{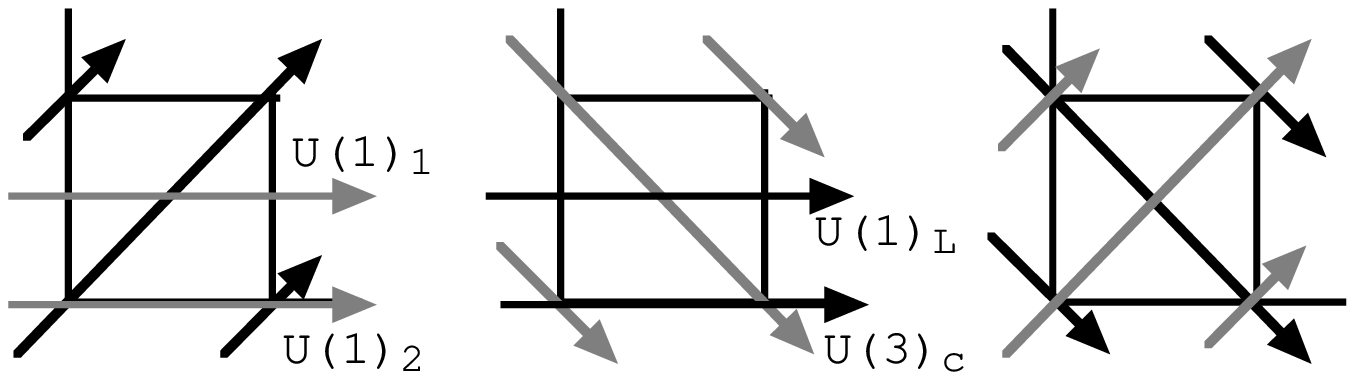,height=40mm}
{\label{D6_2-D6_3}
Actual D6-brane configuration of D6${}_2$ and D6${}_3$ branes.
Gray lines with arrow denote D6-branes of D6${}_3$
 and the solid lines with arrow denote D6-branes of D6${}_2$.
Two stacks of D6-branes of D6${}_3$ in the first torus
 result the difference between the fields with
 weak isospin $1/2$ and $-1/2$.
}

\subsection{Dynamics of the Higgs Sector}
\label{subsec:higgs}

The dynamics of the Higgs sector
 is different from that in other sectors.
The field contents in this sector
 are given in table \ref{contents_higgs}.
\TABLE{
 \begin{tabular}{|c|c|c|}
  \hline
  sector             & $\left(
                         \mbox{SU}(3)_c \times \mbox{SU}(2)_L
                        \right)
                        \times
                        \left(
                         \mbox{USp}(2)_1 \times
                         \mbox{USp}(2)_2 \times
                         \mbox{USp}(2)_3 \times
                         \mbox{USp}(2)_4
                        \right)_{D6_5}$
                     & field \\
                     & ($Y/2, Q_R$)($Q_L, Q_c+Q, Q_1+Q_2$)
                     & \\
  \hline\hline
  $D6_1 \cdot D6_3$  & $(1, 2)(1,1,1,1)_{(+1/2,+1)(-1,0,+1)}
                         \times 2$
                     & $H^{(1)}_i$ \\
                     & $(1, 2)(1,1,1,1)_{(-1/2,-1)(-1,0,+1)}
                         \times 2$
                     & ${\bar H}^{(2)}_i$  \\
  \hline
  $D6_1 \cdot D6_5$  & $(1, 2)(2,1,1,1)_{(0,0)(+1,0,0)}$
                     & $T_a$ \\
                     & $(1, 2)(1,2,1,1)_{(0,0)(+1,0,0)}$
                     & \\
                     & $(1, 2)(1,1,2,1)_{(0,0)(+1,0,0)}$
                     & \\
                     & $(1, 2)(1,1,1,2)_{(0,0)(+1,0,0)}$
                     & \\
  \hline
  $D6_3 \cdot D6_5$  & $(1, 1)(2,1,1,1)_{(+1/2,+1)(0,0,-1)}$
                     & $T^{(+)}_a$ \\
                     & $(1, 1)(1,2,1,1)_{(+1/2,+1)(0,0,-1)}$
                     & \\
                     & $(1, 1)(1,1,2,1)_{(+1/2,+1)(0,0,-1)}$
                     & \\
                     & $(1, 1)(1,1,1,2)_{(+1/2,+1)(0,0,-1)}$
                     & \\
                     & $(1, 1)(2,1,1,1)_{(-1/2,-1)(0,0,-1)}$
                     & $T^{(-)}_a$ \\
                     & $(1, 1)(1,2,1,1)_{(-1/2,-1)(0,0,-1)}$
                     & \\
                     & $(1, 1)(1,1,2,1)_{(-1/2,-1)(0,0,-1)}$
                     & \\
                     & $(1, 1)(1,1,1,2)_{(-1/2,-1)(0,0,-1)}$
                     & \\
  \hline
 \end{tabular}
\caption{
Field contents of the Higgs sector.
Here, $i=1,2$ and $a = 1,2,3,4$.
}
\label{contents_higgs}
}
The dynamics of the gauge interaction
 is that of USp$(2)$ with two massless flavors.
The low-energy effective fields
 after the ``hypercolor'' confinement are
\begin{equation}
 V_a
   = \left[
      \left(
       \begin{array}{c}
        T_a \\ T^{(+)}_a \\ T^{(-)}_a
       \end{array}
      \right)
      \left(
       \begin{array}{ccc}
        T_a & T^{(+)}_a & T^{(-)}_a
       \end{array}
      \right)
          \right]
   = \left(
      \begin{array}{ccc}
       [T_a T_a] & [T_a T^{(+)}_a] & [T_a T^{(-)}_a] \\
                 & [T^{(+)}_a T^{(+)}_a] & [T^{(+)}_a T^{(-)}_a] \\
                 &                       & [T^{(-)}_a T^{(-)}_a] \\
      \end{array}
     \right),
\end{equation}
 where square brackets denote the anti-symmetric contraction
 of the indices of the fundamental representation of
 USp$(2)_a$ in eq.(\ref{gauge_D6_5}).
The field contents after the ``hypercolor'' confinement
 are given in table \ref{contents_higgs_confined}.
\TABLE{
 \begin{tabular}{|c|c|}
  \hline
  field             & $\mbox{SU}(3)_c \times \mbox{SU}(2)_L$ \\
                    & ($Y/2, Q_R$)($Q_L, Q_c+Q, Q_1+Q_2$) \\
  \hline\hline
  $H^{(1)}_i$       & $(1, 2)_{(+1/2,+1)(-1,0,+1)}
                       \times 2$ \\
  ${\bar H}^{(2)}_i$
                    & $(1, 2)_{(-1/2,-1)(-1,0,+1)}
                       \times 2$ \\
  \hline
  $[T_a T_a] \sim S_{H1,a}$
                    & $(1, 1)_{(0,0)(+2,0,0)}
                       \times 4$ \\
  $[T^{(+)}_a T^{(-)}_a] \sim S_{H2,a}$
                    & $(1, 1)_{(0,0)(0,0,-2)}
                       \times 4$ \\
  $[T_a T^{(+)}_a] \sim H^{(2)}_a$
                    & $(1, 2)_{(+1/2,+1)(+1,0,-1)}
                       \times 4$ \\
  $[T_a T^{(-)}_a] \sim {\bar H}^{(1)}_a$
                    & $(1, 2)_{(-1/2,-1)(+1,0,-1)}
                       \times 4$ \\
  \hline
 \end{tabular}
\caption{
Field contents of the Higgs sector
 after the ``hypercolor'' confinement.
Here, $i=1,2$ and $a = 1,2,3,4$.
}
\label{contents_higgs_confined}
}

No superpotential is dynamically generated,
 but the following constraints have to be imposed.
\begin{equation}
 {\rm Pf} V_a = \Lambda_{H,a}^4,
\label{constraint}
\end{equation}
 where $\Lambda_{H,a}$ is the scale of dynamics of USp$(2)_a$.
Since the gauge coupling constants of USp$(2)_a$
 are equal at the string scale,
 the scales of dynamics are also equal:
 $\Lambda_{H,a}=\Lambda_H$.
The constraints of eq.(\ref{constraint})
 mean non-zero vacuum expectation values
 of some low-energy effective fields.
We take the following vacuum expectation values
 which keep maximal symmetry.
\begin{equation}
 \langle V_{a12} V_{a34} \rangle
  = \langle S_{H1,a} S_{H2,a} \rangle = \Lambda_H^4.
\end{equation}
This vacuum expectation values do not break any gauge symmetries
 including anomalous U$(1)$ gauge symmetries,
 because of the moduli dependence of $\Lambda_H$.
If it is possible to factorize the bilinear operator and
\begin{equation}
 \langle V_a \rangle
 = \Lambda_H^2
   \left(
    \begin{array}{cc}
     i \sigma_2 & 0 \\
     0 & i \sigma_2
    \end{array}
   \right),
\end{equation}
 the anomalous U$(1)_L$ and U$(1)_{Q_1+Q_2}$ gauge symmetries
 are spontaneously broken.
Here, we do not discuss this problem in detail,
 and we leave it for future works.

The Yukawa interactions in eq.(\ref{yukawa_higgs})
 give masses to two of four fields of each
 ${\bar H}^{(1)}_a$ and $H^{(2)}_a$.
The values of the masses are determined
 by the values of the Yukawa coupling constants
 and the scale of dynamics $\Lambda_H$.
The composite fields ${\bar H}^{(1)}_1$ and ${\bar H}^{(1)}_3$
 become massive with elementary fields $H^{(1)}_1$ and $H^{(1)}_2$,
 respectively, and
 the composite fields $H^{(2)}_2$ and $H^{(2)}_4$
 become massive with elementary fields
 ${\bar H}^{(2)}_2$ and ${\bar H}^{(2)}_1$, respectively.
Here,
 we are neglecting the contributions
 from the exponentially suppressed Yukawa coupling constants
 as an approximation.
All the massless composite Higgs fields
 ${\bar H}^{(1)}_2$, ${\bar H}^{(1)}_4$, $H^{(2)}_1$ and $H^{(2)}_3$
 have Yukawa interactions with composite quarks and leptons.

\section{Dynamical Generation of Yukawa Coupling Constants}
\label{sec:yukawa}

The higher dimensional interactions of eq.(\ref{to_be_yukawa})
 in superpotential come from the recombination processes
 among open strings at six intersection points:
 (D6${}_2 \cdot$D6${}_4$) - (D6${}_4 \cdot$D6${}_1$)
 - (D6${}_1 \cdot$D6${}_5$) - (D6${}_5 \cdot$D6${}_3$)
 - (D6${}_3 \cdot$D6${}_6$) - (D6${}_6 \cdot$D6${}_2$).
A schematic picture of a recombination process
 is given in figure \ref{recomb}.
\EPSFIGURE{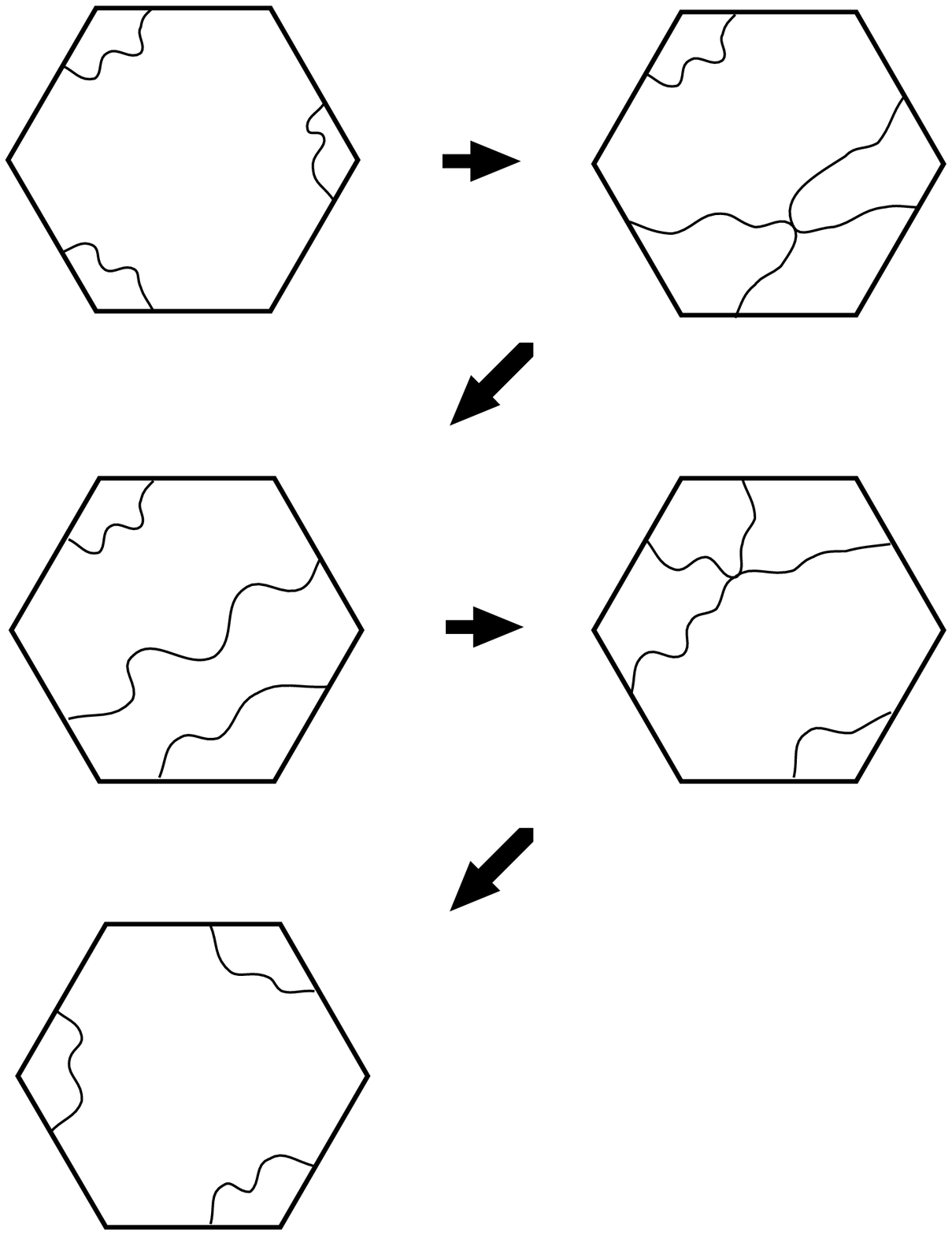,height=9cm}
{\label{recomb}
A schematic picture of a recombination process of open strings.
The apices of the hexagon correspond to six intersection points.
}
These interactions
 give Yukawa interactions for the quark-lepton mass and mixing
 after the ``hypercolor'' confinement.
\begin{eqnarray}
 \sum_{\alpha,\beta=1}^6 \sum_{a=1}^4
 {{g^u_{\alpha \beta a}} \over {M_s^3}}
  [ C_\alpha D_\alpha ] 
  [ {\bar C}_\beta {\bar D}^{(-)}_\beta ]
  [ T_a T^{(+)}_a ]
 &\sim&
 \sum_{\alpha,\beta=1}^6 \sum_{a=1}^4
 g^u_{\alpha \beta a}
 {{\Lambda_L \Lambda_R \Lambda_H} \over {M_s^3}}
 q_\alpha u_\beta H^{(2)}_a,
\\
 \sum_{\alpha,\beta=1}^6 \sum_{a=1}^4
 {{g^d_{\alpha \beta a}} \over {M_s^3}}
  [ C_\alpha D_\alpha ]
  [ {\bar C}_\beta {\bar D}^{(+)}_\beta ]
  [ T_a T^{(-)}_a ]
 &\sim&
 \sum_{\alpha,\beta=1}^6 \sum_{a=1}^4
 g^d_{\alpha \beta a}
 {{\Lambda_L \Lambda_R \Lambda_H} \over {M_s^3}}
 q_\alpha d_\beta {\bar H}^{(1)}_a,
\\
 \sum_{\alpha,\beta=1}^6 \sum_{a=1}^4
 {{g^\nu_{\alpha \beta a}} \over {M_s^3}}
  [ N_\alpha D_\alpha ]
  [ {\bar N}_\beta {\bar D}^{(-)}_\beta ]
  [ T_a T^{(+)}_a ]
 &\sim&
 \sum_{\alpha,\beta=1}^6 \sum_{a=1}^4
 g^\nu_{\alpha \beta a}
 {{\Lambda_L \Lambda_R \Lambda_H} \over {M_s^3}}
 l_\alpha \nu_\beta H^{(2)}_a,
\\
 \sum_{\alpha,\beta=1}^6 \sum_{a=1}^4
 {{g^e_{\alpha \beta a}} \over {M_s^3}}
  [ N_\alpha D_\alpha ]
  [ {\bar N}_\beta {\bar D}^{(+)}_\beta ]
  [ T_a T^{(-)}_a ]
 &\sim&
 \sum_{\alpha,\beta=1}^6 \sum_{a=1}^4
 g^e_{\alpha \beta a}
 {{\Lambda_L \Lambda_R \Lambda_H} \over {M_s^3}}
 l_\alpha e_\beta {\bar H}^{(1)}_a.
\end{eqnarray}
Since the scales $\Lambda_L$, $\Lambda_R$ and $\Lambda_H$
 are the same order of magnitude of $M_s$,
 the low-energy Yukawa interactions
 for the quark-lepton mass are not trivially suppressed.
The values of the elements of the Yukawa coupling matrices of
 $g^u$, $g^d$, $g^\nu$ and $g^e$
 are determined by the D6-brane configuration.
Namely,
 they depend on the places of six intersection points
 in each three torus.
If the separation of intersection points is large,
 the value of the corresponding coupling constant is small,
 and vice versa.
In case that the separation of intersecting points
 is large especially in a large torus,
 the value of the corresponding coupling constant
 is exponentially small.
Therefore,
 we can expect some non-trivial structure
 in Yukawa coupling matrices.

It does not always happen that
 all the six intersection points coincide in all three tori.
Therefore,
 almost all the values of elements in Yukawa coupling matrices
 are small.
This could be a reason
 why the masses of quarks and leptons
 are smaller than the electroweak scale.
Top quark is accidentally heavy in this scenario.

We are not going to explicitly calculate
 the Yukawa coupling matrices.
Instead,
 we describe one specific example
 which gives a quark-lepton mass difference.
Here,
 we investigate the coupling constants
 of the following Yukawa interactions among massless fields.
\begin{equation}
 W_{144} = g^d_{144} q_1 d_4 {\bar H}^{(1)}_4
         + g^e_{144} l_1 e_4 {\bar H}^{(1)}_4.
\end{equation}
The relevant configuration of D6-branes
 is given in figure \ref{yukawa_config}.
\EPSFIGURE{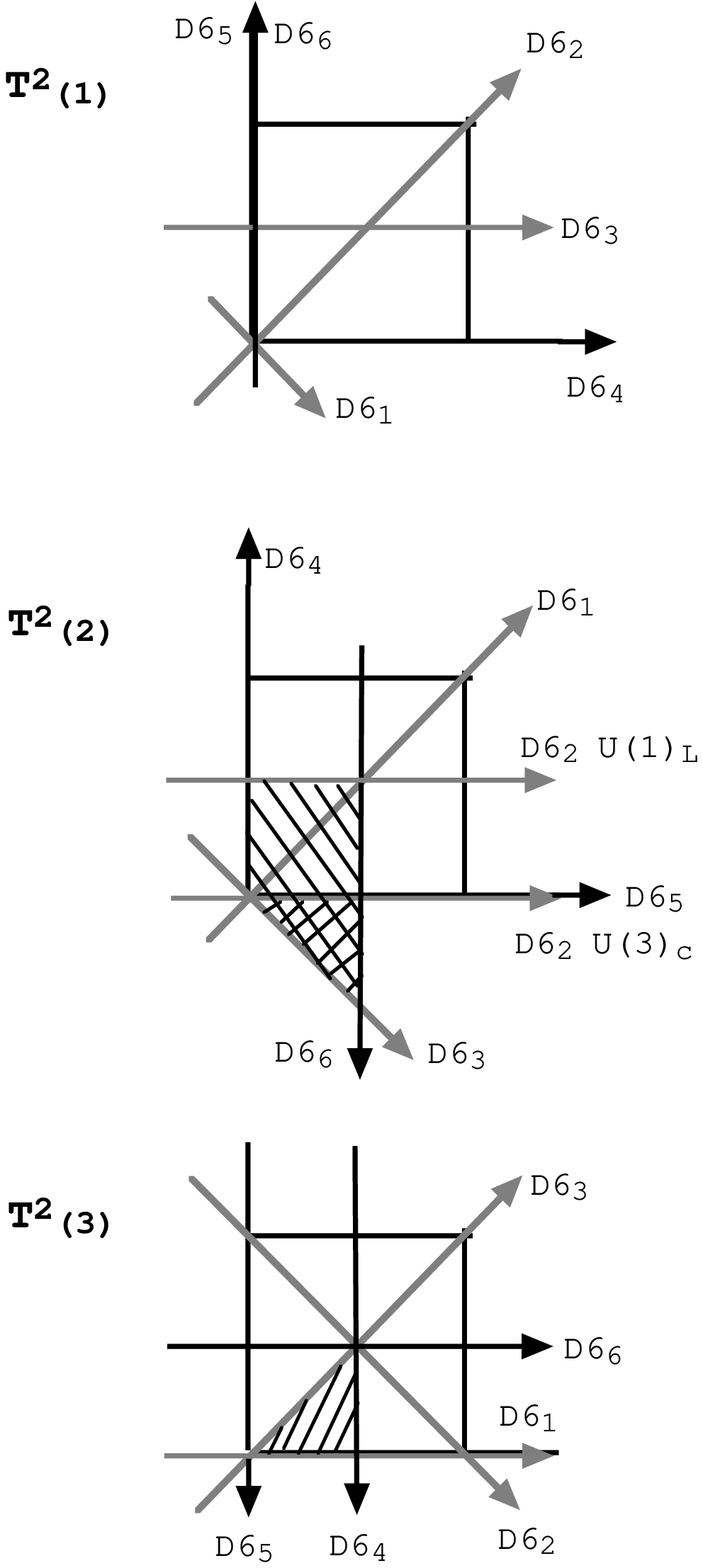,height=16cm}
{\label{yukawa_config}
Configuration of D6-branes
 to investigate of Yukawa coupling constants.
}
In the first torus
 six intersection points do not give finite area,
 and no exponential suppression factor emerges from the first torus.
In the third torus
 six intersecting points are on the apexes of the hatched triangle,
 which results the same exponential suppression factors
 for each Yukawa coupling constant, $g^d_{144}$ and $g^e_{144}$.

The quark-lepton mass difference
 is realized by the two different D6-branes of D6${}_2$
 in the second torus.
The cross-hatched triangular area
 in the second torus in figure \ref{yukawa_config}
 determines an exponential suppression factor to
 the down-type quark Yukawa coupling constant $g^d_{144}$,
 and the hatched quadrilateral area
 (including the cross-hatched triangular area)
 determines another exponential suppression factor to
 the charged lepton Yukawa coupling constant $g^e_{144}$.
Since the quadrilateral area
 is three times larger than the triangular area,
 the charged lepton Yukawa coupling is exponentially smaller
 than the down-type quark Yukawa coupling constant:
 $g^e_{144} \ll g^d_{144}$.
The concept to have hierarchical Yukawa couplings
 utilizing the exponential suppression by the area
 has already been proposed in ref.\cite{AFIRU-2}.
The above scenario is a non-trivial realization of this concept.

To investigate the Yukawa coupling matrices further,
 we have to treat the problem much more precisely.
The mass mixings in
 the left-handed sector, right-handed sector and Higgs sector
 should be solved including exponentially suppressed
 Yukawa interactions among ``preons'' at the string level.
The normalization of
 composite fields in the K\"ahler potential
 (the normalization of the kinetic terms of composite fields)
 should also be considered.
The full structure of Yukawa coupling matrices
 should be investigated in some specific D6-brane configurations
 which give some intuitive idea to explain
 the realistic quark-lepton mass spectrum and flavor mixings.

The electroweak symmetry breaking
 by the vacuum expectation values of Higgs fields
 is required to generate the masses of quarks and leptons
 as well as weak bosons.
Although supersymmetry breaking is necessary
 for non-zero vacuum expectation values of Higgs fields
 in this model
 (of course, supersymmetry breaking is also required
  to give masses to superpartners),
 there are no explicit mechanism of supersymmetry breaking
 in this model.
There is no hidden sector
 which can be realized by some D6-brane with no intersection
 with any other D6-branes.
If there is a hidden sector D6-brane
 on which super-Yang-Mills theory is realized,
 supersymmetry could be dynamically broken
 by gaugino condensation with supergravity effects.
If gauginos can condense
 in super-Yang-Mills theories with massless matter
 as assumed in ref.\cite{CLW}
 (this means supersymmetry breaking
 through the Konishi anomaly relation without supergravity effects),
 there is a possibility of the dynamical supersymmetry breaking
 by the ``hypercolor'' dynamics in this model.

\section{Summary and Conclusions}
\label{sec:conclusions}

We have constructed a supersymmetric composite model
 in type IIA ${\bf T^6}/({\bf Z_2} \times {\bf Z_2})$ orientifolds
 with intersecting D6-branes.
Four generations of quarks and leptons
 are composite particles in this model.
Two pairs of electroweak Higgs fields are also composite.
The expected mechanisms to give masses to exotic particles
 should be further investigated.
There is an additional non-anomalous U$(1)$ gauge symmetry
 which should be spontaneously broken above the electroweak scale.
This model is a toy model to illustrate
 a new mechanism of dynamical generation of Yukawa couplings
 for the masses and mixings of quarks and leptons.

The Yukawa interactions for the quark-lepton mass
 can be naturally generated
 by the interplay between higher dimensional interactions
 at the string level
 (six-dimensional operators in the superpotential
  in the four-dimensional effective theory)
 and ``hypercolor'' dynamics.
The actual value of the Yukawa coupling constants
 are determined by the D6-brane configuration in three tori.
Although the full structure of the Yukawa coupling matrices 
 have not yet been investigated,
 a general mechanism to have small Yukawa coupling constants
 and a specific example which gives quark-lepton mass differences
 have been introduced.

There is no hidden sector in this model,
 and the explicit mechanism of supersymmetry breaking is absent.
Although the ``hypercolor'' dynamics might break supersymmetry,
 the concrete analysis of the dynamics and
 the mediation of supersymmetry breaking
 is left for future works.

Introduction of the compositeness of
 Higgs fields and/or quark and lepton fields
 is one of the natural directions in the model building
 with intersecting D-branes.
Many additional D-brane are required in addition to
 the D-branes for the gauge symmetry of the standard model
 to satisfy the tadpole cancellation conditions.
It is natural to consider that
 these additional D-branes act some important roles in Nature.

It would be very interesting
 to explorer more realistic model in this framework.

\end{document}